\documentclass[sigplan,nonacm]{acmart}

\usepackage{graphicx} 
\usepackage{mathtools}
\usepackage{mathpartir}
\usepackage{amsfonts}
\usepackage{amsmath}
\usepackage{stmaryrd}
\usepackage{amsthm}
\usepackage{listings}
\usepackage{subcaption}
\usepackage{microtype}

\usepackage[nofillcomment,linesnumbered,ruled,vlined,noend]{algorithm2e} 

\SetCommentSty{mycommfont} 

\usepackage{cleveref}

\definecolor{todocolor}{rgb}{0.8,0,0}
\definecolor{editcolor}{rgb}{0,0,0.8}
\newcommand{\TODO}[1]{{\color{todocolor}#1}}
\newcommand{\IGNORE}[1]{}

\renewcommand{\infer}[3][]{
\ifthenelse{\equal{#1}{}}{
\inferrule{#2}{#3}
}
{
\inferrule*[right={\scriptsize \textbf{#1}}]
{#2}
{#3}
}
}

\definecolor{keywordcolor}{rgb}{0.5,0,0.5}
\definecolor{textgray}{gray}{0.4}
\definecolor{mygray}{rgb}{0.5,0.5,0.5}
\title{On the Duality of Task and Actor Programming Models}

\author{Rohan Yadav}
\affiliation{%
    \institution{Stanford University}
    \country{USA}
}
\email{rohany@cs.stanford.edu}
\author{Joseph Guman}
\affiliation{%
    \institution{NVIDIA}
    \country{USA}
}
\email{joeytg@cs.stanford.edu}
\authornote{This work was done while the author was at Stanford.}
\author{Sean Treichler}
\affiliation{%
    \institution{NVIDIA}
    \country{USA}
}
\email{sean@nvidia.com}
\author{Michael Garland}
\affiliation{%
    \institution{NVIDIA}
    \country{USA}
}
\email{mgarland@nvidia.com}
\author{Alex Aiken}
\affiliation{%
    \institution{Stanford University}
    \country{USA}
}
\email{aiken@cs.stanford.edu}
\author{Fredrik Kjolstad}
\affiliation{%
    \institution{Stanford University}
    \country{USA}
}
\email{kjolstad@cs.stanford.edu}
\author{Michael Bauer}
\affiliation{%
    \institution{NVIDIA}
    \country{USA}
}
\email{mbauer@nvidia.com}

\begin{document}

\begin{abstract}

%
%
%
%
%
%
%
%
Programming models for distributed and heterogeneous machines are rapidly growing in popularity to meet the demands of modern workloads.
Task and actor models are common choices that offer different trade-offs between development productivity and achieved performance. 
Task-based models offer better productivity and composition of software, whereas actor-based models routinely deliver better peak performance due to lower overheads.
While task-based and actor-based models appear to be different superficially,
we demonstrate these programming models are duals of each other.
%
Importantly, we show that this duality extends beyond functionality to performance, and elucidate techniques that let task-based systems deliver performance competitive with actor-based systems without compromising productivity.
We apply these techniques to both Realm, an explicitly parallel task-based runtime, as well as Legion, an implicitly parallel task-based runtime.  
%
We show these techniques reduce Realm's overheads by between 1.7-5.3x,
coming within a factor of two of the overheads imposed by
heavily optimized actor-based systems like Charm++ and MPI.
%
We further show that our techniques enable
between 1.3-5.0x improved strong scaling of unmodified Legion applications.
%

%
%

\end{abstract}

\maketitle

\section{Introduction}

Modern workloads for distributed and heterogeneous machines place stringent demands on programming systems, which must deliver both high productivity to facilitate rapid program evolution and low overhead to extract maximum performance from hardware.
%
In an attempt to meet these demands, both task-based~\cite{legion, realm, regent, parsec, starpu, ray} and actor-based~\cite{charm++, mpi, ray, cpp-native-actors, orleans, erlang} systems have been adopted for many important applications.
Unfortunately, existing implementations of task-based and actor-based systems fail to completely deliver both high productivity and performance. 
By understanding the nature of the compromises that clients must make when choosing either an actor- or a task-based system, we can uncover the deep relationship between the programming models and eliminate the need for compromise
for an important class of applications.

Actor-based programming models are the basis for some of the first systems for distributed memory machines~\cite{mpi, charm++, erlang}.
Actor models send and receive explicit messages between \emph{actors} (either objects or processes) to perform data movement and synchronization.
Due to the simplicity of the message passing interface in most actor models, the underlying systems are embodied by thoroughly optimized implementations to minimize overheads associated with sending and receiving messages.
While the simplicity of actor models ensures low-overhead implementations, it often incurs a latent cost: as programs become larger and more complex, the burden of maintaining their correctness tends to scale super-linearly. 
Sophisticated programs accumulate interacting features that actors must support with a burgeoning set of asynchronous messages that may arrive in a growing set of permutations.
In the worst case, the number of permutations that must be handled grows factorially with the total number of message types, requiring a complex state machine in each actor.
Consequently, actor models have been subject to the criticism that they are error-prone and result in programs that are difficult to maintain and evolve~\cite{spinifel-experience}.

In contrast, task-based programming models strive to deliver higher productivity in response to the increasing complexity of modern hardware. 
Programs are organized as a directed acyclic graph (DAG), often constructed dynamically,
of short-lived computations called \emph{tasks}.
%
Explicitly-parallel task-based models require clients to directly construct the DAG by specifying data movement and dependencies between tasks, while implicitly-parallel models infer the DAG from the data usage of tasks.
Regardless of the DAG construction mode, task-based programs are simpler to modify with only local reasoning, making it easier to compose modules together and maintain software over long spans of time~\cite{spinifel-experience}.

Due to the generality of the DAG execution model, task-based systems frequently suffer from higher overheads associated with scheduling and executing the DAG of tasks.
If the granularity of tasks is sufficiently large, these overheads have a negligible impact on performance.
However, under strong-scaling conditions or in applications with fine-grained parallelism, the overheads
can eventually inhibit performance as they come to dominate the runtime of the program.
Modern accelerators, with growing compute power, are progressively shrinking the execution times for tasks with each new generation, thereby placing increasing pressure on task-based systems to lower their overheads to maintain scalability.  
The complementary nature of the trade-offs associated with tasks and actors suggests that there exists a deeper relationship between the two classes of programming models.
Inspired by the classic duality between message-based and procedure-based operating systems~\cite{os-duality}, we make the observation that actor and task programming models are also duals of each other.
Actor-based programs are characterized by long-running actors that communicate through messages, similar to processes in message-based operating systems.
%
In contrast, task-based programs are characterized by short-lived tasks that operate on common data structures similar to the procedures in procedure-based operating systems.
%
Importantly, we further claim that this duality extends beyond functional equivalence, and is additionally a performance duality where programs written in one style can be rewritten in the other and achieve comparable performance. 

\begin{figure}
    \centering
    \includegraphics[width=0.9\linewidth]{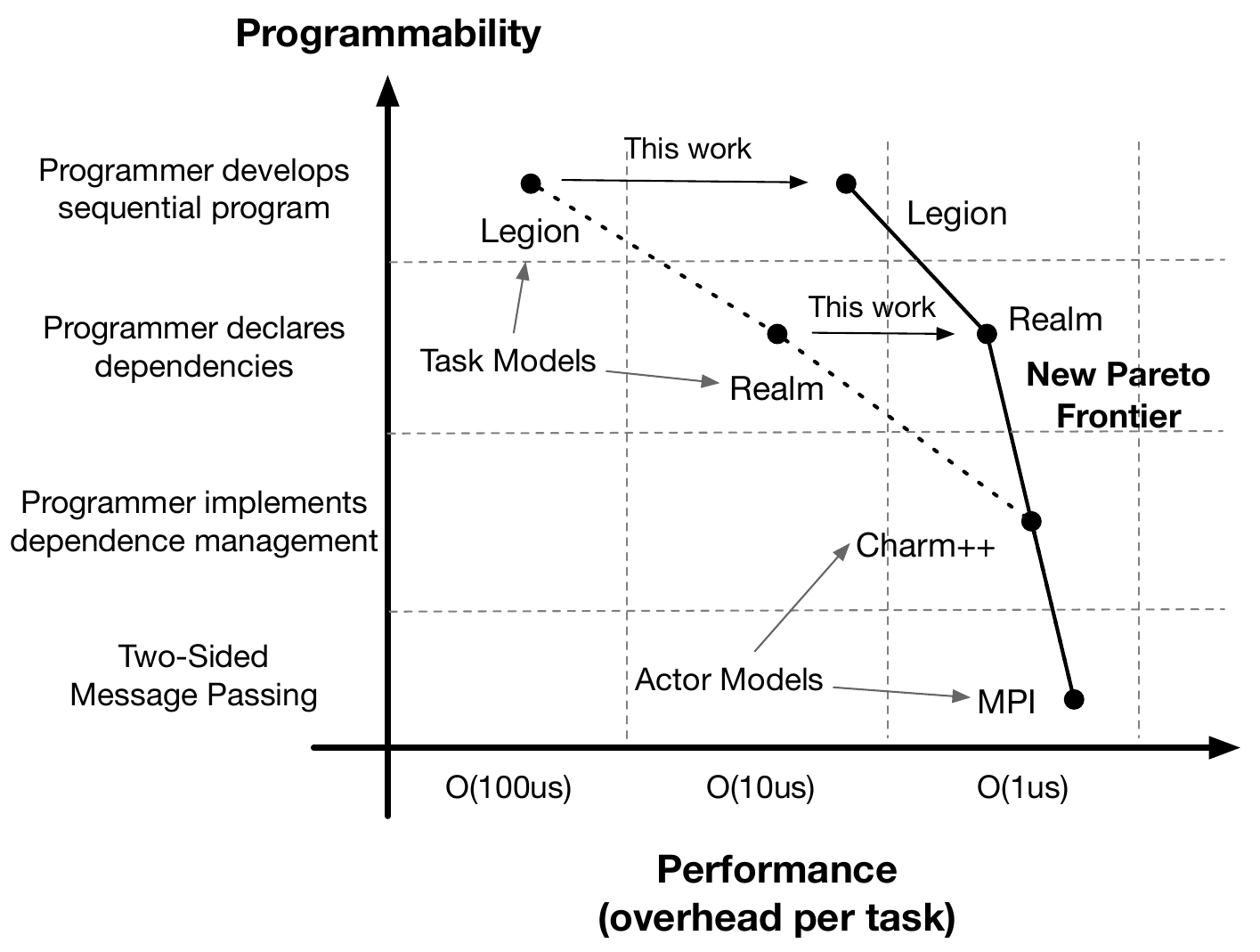}
    \caption{This work defines a new Pareto frontier for distributed
    programming along the performance and programmability tradeoff. See \Cref{sec:evaluation} for performance details.}
    \label{fig:pareto-frontier}
\end{figure}

We illuminate the sources of overheads imposed by task-based models
through different translation strategies between task-based and
actor-based models.
We then leverage these insights to develop compilation techniques for task-based
programming systems that translate demarcated subgraphs of the complete program
DAG into actor-based programs.
The compilation strategy results in a set of surprisingly simple actors that efficiently
execute the target subgraph, greatly reducing overheads for important and repeatedly
executed components of the complete program DAG.
Iterative applications, such as those present in domains such as deep learning and scientific
computing, are amenable to our graph compilation techniques.
As depicted in \Cref{fig:pareto-frontier}, our work exploits the actor-task duality 
in the task-to-actor direction to bridge the gap between the performance of task-based models
and actor-based models while preserving their programmability characteristics.


We implement our techniques within the explicitly-parallel
task-based system Realm~\cite{realm}, introducing
a subgraph compilation module.
%
We modify the implicitly-parallel task-based system Legion~\cite{legion}
to target this compilation module when memoizing its dynamic 
analysis~\cite{legion-tracing, auto-tracing}, automatically improving performance.
%
Our work is applicable beyond Realm and Legion to improve the performance
of both explicitly-parallel and implicitly-parallel task-based systems.

The specific contributions of this work are:
\begin{enumerate}
    \item An exploration of the duality and trade-off space between actor- and task-based
    programming models.
    \item A compilation strategy that decreases the overheads
    of both explicitly-parallel and implicitly-parallel task-based
    programming models.
    
\end{enumerate}

To evaluate our work, we measure the performance of Legion and Realm
within the Task Bench~\cite{task-bench} framework.
We show that our techniques reduce the smallest task granularity
efficiently supported by both systems (see \Cref{sec:eval-task-bench})
by 3.3x-7.1x and 1.7x-5.3x respectively.
These optimizations allow Realm to come within a factor of two of 
actor-based models like Charm++ and MPI, which (to the best of our knowledge)
has not yet been demonstrated by any existing task-based runtime systems.
We then show that our techniques improve the strong-scaling
performance of unmodified Legion applications by between 1.3x-5.0x.


\section{Background}


\subsection{Actor-Based Programming Models}

Actor-based models are characterized by
long-lived, stateful objects called \emph{actors} that maintain
arbitrary local state and communicate through asynchronous
messages.
In modern actor-based systems, actors are often
associated with 
machine resources (e.g., a GPU or a CPU core),
and applications are comprised of
actors performing local computations and notifying other
actors to start follow-up work.

Actor-based programming models are well-studied and have been 
extensively formalized in prior work~\cite{agha-actors, actors-grief}.
For simplicity and the focus of this work, we consider
a simple actor-based runtime that might be embedded within
a standard host language, as shown in \Cref{fig:actor-based-runtime-api}.
This runtime is a simple model of systems
like Charm++~\cite{charm,charm++} or Ray~\cite{ray}.
The \texttt{Actor} object is extended (in potentially multiple ways) by the
application to contain arbitrary private state and an implementation 
of the \texttt{handle\_message} function.
Actors are registered with the runtime to a concrete resource.
%
%
The main primitive offered by the runtime
is the ability to send a target actor a message, which (remotely) invokes
the target actor's \texttt{handle\_message}.
Applications in actor-based programming models are often structured as state machines
(discussed more in \Cref{sec:specialized-actors}) that accept messages until
an expected set is received to execute some application computation.
These state machines only modify the private state of the actor handling each message;
all inter-actor communication and coordination occurs through message passing.

While not a common classification, we also consider two-sided messaging systems like MPI
to be within the actor-based programming model family, as they share the properties
of having long-running processes that react to incoming messages.
%
%
Two-sided messaging can be simulated with the actor model we present by
buffering arriving messages until they are handled by the corresponding receive operations (which is how MPI implements non-blocking messages).

\subsection{Task-Based Programming Models}

Task-based programming models emerged after actor-based
programming systems, aiming to provide more composable and more
accelerator-friendly abstractions for parallel computing.
The core concept in task-based programming models is a
\emph{task}, which is a stateless, user-defined function.
Tasks may launch other tasks, and are issued onto a target
processor to run asynchronously from the launching task.
Task-based applications express their computation as a
graph of tasks that operate over shared data structures.
These task graphs can be constructed in an offline or static
manner~\cite{parsec-ptg}, or in an online, dynamic
fashion~\cite{realm, legion, starpu, ray, parsec}.

\begin{figure}
\begin{lstlisting}
class Processor;
class ActorRT {
  void send_message(int aid, int mid, void* args, int len);
  void register_actor(Actor* a, int aid, Processor target); };
// Actor is extended by the application.
class Actor {
  // Actors maintain arbitrary, but private state.
  void handle_message(int mid, ActorRT* rt, void* args); };
\end{lstlisting}
\caption{Actor-based Runtime System}
\label{fig:actor-based-runtime-api}
\end{figure}

\begin{figure}
\begin{lstlisting}
class Event, Allocation, Processor, Memory;
using Events = vector<Event>;
class TaskRT {
  void register_task(int tid, void (*task) (TaskRT*, void*));
  Event launch(Processor p, int tid, void* args, Events pre);
  pair<Event, Allocation> alloc(Memory m, int s, Events pre);
  Event copy(Allocation src, Allocation dst, Events pre); };
void task(TaskRT*, void* args) { /* Stateless task body. */ }
\end{lstlisting}
\caption{Task-based Runtime System}
\label{fig:task-based-runtime-api}
\end{figure}

\Cref{fig:task-based-runtime-api} presents a simplified interface
for an explicitly-parallel task-based system that supports
dynamic task graph construction (like Realm~\cite{realm}).
All operations in the task-based model return an \emph{event}
that represents the asynchronous completion
of the operation.
%
Applications may launch tasks on
processors, allocate data in memories across the
machine, and copy data between allocations.
Unlike actor-based models, tasks may have side-effects on
allocations that outlive their individual lifetimes, and tasks can
share state through these side-effects.
Each asynchronous operation is predicated on a set of events that
must complete before the operation executes.
The task-based runtime schedules operations that have all
event preconditions satisfied, automatically overlapping the execution
of independent operations (such as data movement and computation).
%
%
%
The interface in \Cref{fig:task-based-runtime-api} can be embedded within
a general purpose language, and is the target for arbitrary
computation to dynamically construct a task graph by computing dependencies and
issuing tasks.

\Cref{fig:task-based-runtime-api}  models an explicitly-parallel
task-based system, where the programmer is responsible for specifying 
dependencies (represented with events)
between computations.
Implicitly-parallel task-based systems, such as Legion~\cite{legion} or StarPU~\cite{starpu},
leverage a higher-level program representation where tasks describe what data
they will access, and then the system performs an analysis to discover the necessary
dependencies between issued tasks.
These higher-level systems can express the resulting task graph after the dependence
analysis using an explicitly-parallel task-based model.
For the discussions of duality and compilation in Sections~\ref{sec:duality} and~\ref{sec:compilation},
when referring to task-based models, we are specifically discussing explicitly-parallel
models; we return to implicitly-parallel models in \Cref{sec:implicit-par}.

\section{Equivalence and Duality}\label{sec:duality}

We now describe a functional equivalence and performance duality between
actor-based and task-based models.
We are inspired by the work of Lauer et al.~\cite{os-duality},
which showed an equivalence and duality between message- and procedure-based
operating systems.
The duality arises in how applications developed in
either system can share core application logic, while the differences
arise only in the synchronization of when that shared application logic should execute.
We reveal this duality through different reduction strategies that expose the structure underlying
the two programming models.
We then discuss the inherent tradeoffs made between performance and programmability
in the two models.

\subsection{Actors $\rightarrow$ Tasks}

We reduce actors to tasks by demonstrating
an actor-based system implemented with a task-based system.
%
While tasks are stateless functions, tasks may emulate stateful actors 
by providing them access to persistent state.
The reduction is straightforward and shown in \Cref{fig:actor-to-task-reduction}.
%
%
%
Notably, the reduction does not leverage the task-based model's event/dependence infrastructure
and is relatively opaque; coordination and communication is still the responsibility of the programmer, 
hidden inside the existing actors' message handler implementations.
However, this reduction can preserve performance by eschewing the dependence infrastructure.
Each task launch with no preconditions can be implemented efficiently with a single message, similar
to the execution of the actor program without the reduction.
While simple, this reduction also models the core of systems like Ray~\cite{ray},
which provide actors and tasks in the same language.

\begin{figure}
\begin{lstlisting}
// Let A be an actor handling messages M1 and M2, registered
// to processor P. Create an allocation for A in a memory
// visible to P. For simplicity, ignore the returned event.
auto result = task_rt->alloc(P.memory(), sizeof(A), {});
A* stateA = result.second.get_base_pointer();
// Register tasks for each message A handles.
void taskAM1(void* args) { stateA->handle_message(M1, args); }
void taskAM2(void* args) { stateA->handle_message(M2, args); }
register_tasks({A_M1, task_A_m1}, {A_M2, task_A_m2});
// actor_rt->send_message(A, args) translates to a task launch
// with no event preconditions, which can be invoked from
// anywhere on the machine.
task_rt->launch(P, A_M1, args, {});
\end{lstlisting}
\caption{Reduction pseudocode from actors to tasks.}
\label{fig:actor-to-task-reduction}
\end{figure}

\begin{figure}
\begin{lstlisting}
// A simple runtime system creates one worker per processor
// and a scheduler that interfaces with the application.
class Scheduler : Actor {
  void handle_message(int mid, void* args) {
    switch (mid) {
      case LAUNCH: {
        auto [tid, p, ev, targs, preds] = unpack(args);
        register_pending_task(ev, p, preds, {tid, targs});
        send_message(this, SCHEDULE, {}); }
      case TASK_DONE: {
        Event ev = unpack(args);
        notify_waiting_tasks(ev);
        send_message(this, SCHEDULE, {}); }
      case SCHEDULE: {
        for (auto [ev, tid, p, args] : get_ready_tasks()) {
          send_message(get_worker(p), EXEC, {ev, tid, args});
        }}}}}
class Worker : Actor {
  void handle_message(int mid, void* args) {
    assert(mid == EXEC);
    auto [tid, ev, targs] = unpack(args);
    execute_task_body(tid, task_args);
    send_message(SchedulerID, TASK_DONE, {ev}); }}
// Tasks are launched by sending a message to the scheduler.
Event launch(Processor p, int tid, void* args, Events pre) {
  Event ev = generate_fresh_event();
  send_message(get_sched(), LAUNCH, {tid, p, ev, args, pre});
  return ev; }
\end{lstlisting}
\caption{Runtime reduction pseudocode for tasks to actors.}
\label{fig:task-to-rt-actor-reduction}
\end{figure}

\begin{figure*}

    
    

    


\begin{subfigure}[b]{0.33\textwidth}
\centering
\includegraphics[width=0.45\textwidth]{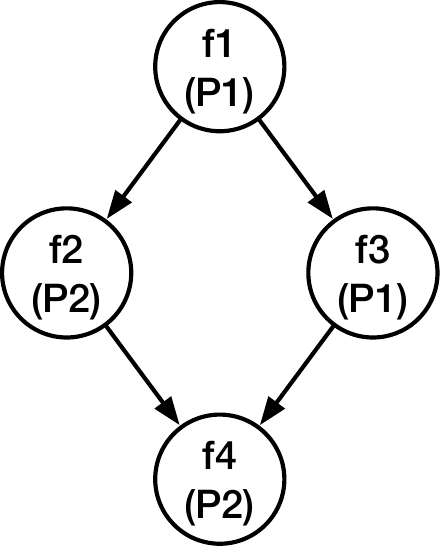}
\caption{Example program DAG.}
\label{fig:simple-dag}
\end{subfigure}%
\begin{subfigure}[b]{0.66\textwidth}
\centering
\begin{minipage}{0.49\textwidth}
\begin{lstlisting}
class P1Actor : Actor {
  void handle_message(int id, void*) {
    if (id == F1_START) {
      f1();
      send_message(P2_ACTOR, F2_START);
      // Use a message send to make
      // the F3 start state explicit.
      send_message(P1_ACTOR_ID, F3_START);
    } else if (id == F3_START) {
      f3();
      send_message(P2_ACTOR_ID, F4_START);
    }}};
\end{lstlisting}
\end{minipage}\hfill%
\begin{minipage}{0.49\textwidth}
\begin{lstlisting}[firstnumber=last]
class P2Actor : Actor {
  // Maintain an notification count
  // for the F4 start state.
  int count = 2;
  void handle_message(int id, void*) {
    if (id == F2_START) {
      f2();
      send_message(P2_ACTOR_ID, F4_START);
    } else if (id == F4_START) {
      if (atomicSub(&count, 1) == 0)
        f4();
    }}};
\end{lstlisting}
\end{minipage}
\caption{Actor-based implementation of \Cref{fig:simple-dag}.}
\label{fig:simple-dag-actors}
\end{subfigure}

\begin{subfigure}[b]{0.4\textwidth}
\centering
\begin{lstlisting}
register_tasks({TID_F1, f1}, {TID_F2, f2},
               {TID_F3, f3}, {TID_F4, f4});
Event e1 = launch(P1, TID_F1);
Event e2 = launch(P2, TID_F2, e1);
Event e3 = launch(P1, TID_F3, e1);
Event e4 = launch(P2, TID_F4, {e2, e3});
\end{lstlisting}
\caption{Task-based implementation of \Cref{fig:simple-dag}.}
\label{fig:simple-dag-tasks}
\end{subfigure}%
\begin{subfigure}[b]{0.6\textwidth}
\centering
\includegraphics[width=0.8\textwidth]{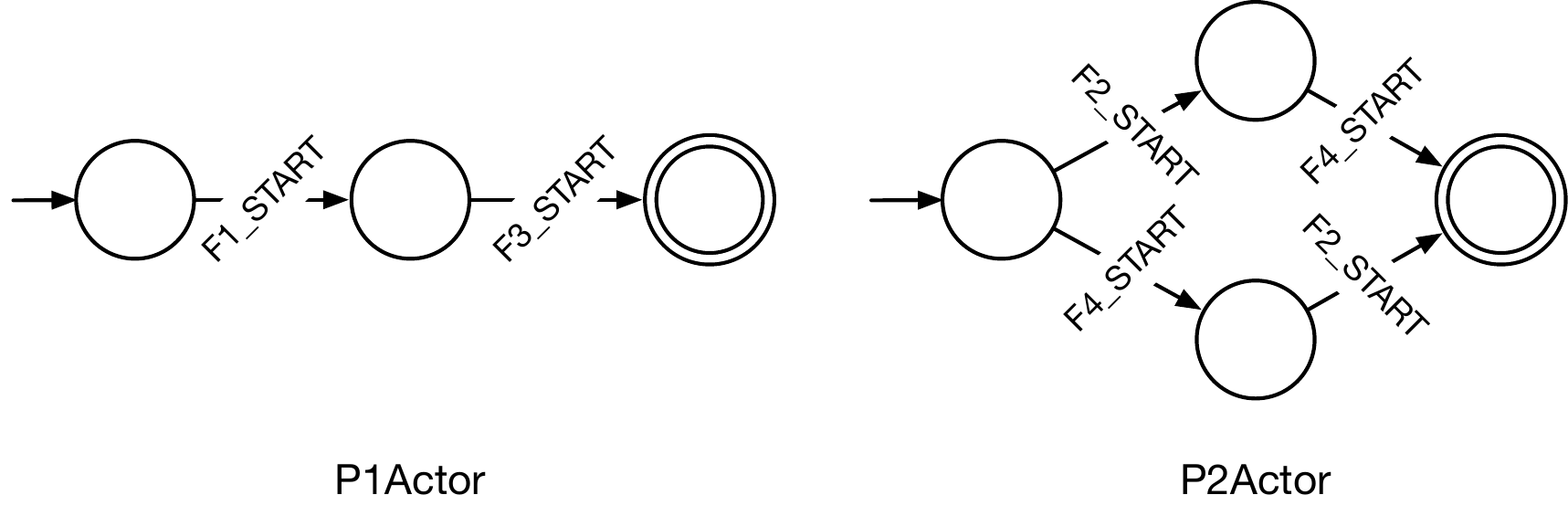}
\caption{Visualization of actor state machines.}
\label{fig:simple-dag-actor-state-machine}
\end{subfigure}

\caption{Example computation developed in actor-based and task-based programming models.}
\label{fig:simple-dag-all}
\end{figure*}

\subsection{Tasks $\rightarrow$ Actors}\label{sec:generic-task-actor-reduction}

We now reduce task-based models onto actor-based models,
which starts to expose the structure of the programming model design space.
We discuss two fundamentally different reduction strategies, which
encapsulate different points in a tradeoff between performance
and programmability.


The first strategy is to construct a collection of actors that form
a \emph{runtime system} to execute an \emph{arbitrary} DAG that
is constructed \emph{dynamically} and \emph{incrementally} as tasks
are issued into the system one at a time.
A simple set of actors structured like a runtime system are described
in \Cref{fig:task-to-rt-actor-reduction}, which contains a \emph{scheduler}
actor that manages pending tasks and event dependencies along with
a set of \emph{worker} actors that execute ready tasks on different processors.
%
This reduction strategy is a simplified model of how task-based systems
are implemented today, as independent processes communicating through
active messages or remote procedure calls.
%

This reduction establishes functional equivalence between
actor-based and task-based models, but also reveals where
the performance differences between the two models arise.
%
Task-based models allow for the declarative specification of
dependencies between computations, but this abstraction comes at the
cost of a generic runtime system.
In contrast, high-performance actor-based applications avoid generic
dependence infrastructure and synchronize in an \emph{application-specific} manner.
Actors directly encode the specific communication patterns and dependence logic for a particular application, rather than through an intermediate layer. 

\subsection{Specialization in Actor Models}\label{sec:specialized-actors}


To gain intuition for the specialization in
efficient actor programs, we develop an example in \Cref{fig:simple-dag-all}.
Any parallel computation can be described as a directed acyclic graph (DAG), where
the vertices represent atomic computations, and the edges represent dependencies
between computations.
The DAG itself may be dynamically constructed, and represents
an ``unrolled'' version of the application.
\Cref{fig:simple-dag} depicts a DAG with the application logic contained
in the functions \texttt{f1}, \texttt{f2}, \texttt{f3} and \texttt{f4}, which
are specified to be executed on the processors \texttt{P1} and \texttt{P2}.
Actor-based and task-based implementations of this DAG are shown in \Cref{fig:simple-dag-actors}
and \Cref{fig:simple-dag-tasks} respectively.
%
The task-based implementation wraps the application logic in tasks and directly
translates the DAG into tasks and events.
In contrast, actor-based programs are frequently structured as
\emph{state machines} that directly communicate to negotiate
dependencies, instead of through an intermediate runtime layer.
Specializing the actors to a specific DAG avoids overheads that plague
task-based systems 
such as extra coordination messages, dynamic dependence management, and task
submission costs.

In the general case, a task-based system must be implemented in the generic flavor
discussed in \Cref{sec:generic-task-actor-reduction}, as the desired DAG
is expressed in a dynamic and incremental manner: tasks may be spawned at any time from any processor,
and may depend on an event produced anywhere in the distributed system.
However, if the unit of work submission is a DAG instead of a single task,
there is an opportunity to perform the same specializations that users of actor-based
models employ, at least within that DAG.
In particular, a \emph{fixed} DAG $G$ of tasks can be compiled into
a set of actors that directly communicate and eschew the standard
dependence infrastructure of the task-based runtime system,
sending exactly one cross-processor message per edge in $G$.
We present an algorithm to perform this specialization in \Cref{sec:compilation},
and demonstrate that an implementation  in the Realm runtime system lowers overheads
by up to a factor of five.

\subsection{Duality and Tradeoffs}

Actor-based and task-based models serve as different vehicles for programmers 
to materialize a DAG into a concrete implementation.
The atomic computations represented by the nodes of the DAG may be abstracted from the
implementation substrate, as done in \Cref{fig:simple-dag-all},
and be used in either a task-based or actor-based implementation.
%
The remaining difference between
programs in the two models is the implementation of dependence management
that coordinates the shared application logic; that responsibility is either
the programmer's (actors) or the runtime system's (tasks).
This glue code (state machine construction or declarative task-graph construction) 
is separate from the shared application code, and we have shown in the previous
sections how either construction may be translated into the other.

While Lauer et al.~\cite{os-duality} conclude that the performance of
message-based and procedure-based operating systems is dependent on the
underlying hardware, we believe that the choice of actors versus tasks
is a tradeoff between performance and productivity.
Actor-based models have been historically viewed and experimentally
shown~\cite{task-bench} to impose significantly lower overheads than task-based
models, due to the specialization discussed in \Cref{sec:specialized-actors}.
We argue that this performance comes at a cost of higher burden on the programmer
to manage the dependencies between computations and to exploit
available parallelism.

%
As seen in \Cref{fig:simple-dag-actor-state-machine}, 
actor state machines must 
explicitly manage communication and  
ensure all dependencies are correctly met all while
simultaneously exploiting as 
much parallelism as possible.
When the application DAG is modified to include a new vertex or edge,
or moves a computation onto a new processor, the
state machines may require significant modifications.
These alterations may include new messages and communication patterns,
changes to the expected number of incoming messages for a state,
or new states to handle the interaction between
new vertices and all existing vertices that may potentially execute in parallel.
Targeting accelerators like GPUs incurs additional 
complexity to actor state machines.
A common paradigm in actor models~\cite{charm++} is to treat
the completion of asynchronous accelerator computations as
additional messages sent to an actor, increasing the number
of states and messages to reason about.
%

In contrast, task-based models specify dependencies
in a declarative style and leave satisfying those
dependencies while maximizing parallelism to the runtime system.
This declarative nature means that program modifications like adding new dependencies,
new tasks, or changing where tasks run are considerably easier.
Additionally, the declarative specification naturally incorporates asynchronous accelerators --- the runtime is responsible for ensuring that tasks dependent
on asynchronous work start only when the work completes.
Higher-level, implicitly parallel task-based systems \cite{legion, starpu}
further simplify programming by automatically inferring parallelism, not even 
requiring the user to describe dependencies~\cite{spinifel-experience}.
As discussed, task-based systems have historically traded this ease of 
programming for overheads that come from implementing the abstractions they
provide.
For example, the task-based Realm system has up to 7 times the overhead
of an efficient actor system like Charm++ (\Cref{sec:evaluation}).
In this work, we make significant progress towards collapsing this tradeoff space for
applications that are able to present repeatedly executed subgraphs to the
task-based runtime system, such as those present in iterative computations that are found in domains spanning the gamut from deep learning to scientific computing.
For such applications, we are able to provide the programmability
properties of task-based programming models while delivering overheads
approaching efficient actor-based programming models.

%
%

\section{Compiling Task Graphs to Actors}\label{sec:compilation}

We now leverage the duality between actors and tasks to
develop a compilation strategy that reduces the overheads
imposed by task-based models.
Our approach lowers task graphs to a set of
specialized actors that avoid dynamic dependence
management and extraneous communication.
%
%
We then show in \Cref{sec:implicit-par} how to leverage this intermediate
representation from an implicitly-parallel task-based system.

\subsection{Interface}


Applications define a task graph $G$ with a set of vertices
$V$ and edges $E$.
Each vertex in $V$ is an operation, such as a task
or copy, and edges describe dependencies between operations.
Vertices may also be \emph{external} pre- or post-conditions,
representing dependencies that either come from outside of the
graph, or must be notified when operations within
the graph complete.
Applications dynamically construct graphs and register them
with the runtime; afterwards, the entire graph may be
launched as a single operation.
The target graphs for compilation are subgraphs of the complete
program graph that are performance critical and repeatedly executed.
The compiled subgraphs are then dynamically issued by the application and stitched
into the larger program graph through the external pre- and post-conditions.
Control structures such as loops and conditionals are the purview of the
application dynamically building the larger program graph, rather than constructs 
within the compiled subgraph.






\subsection{Compilation}

\begin{figure*}[t]
\begin{minipage}{0.63\textwidth}
\centering
\includegraphics[width=0.85\textwidth]{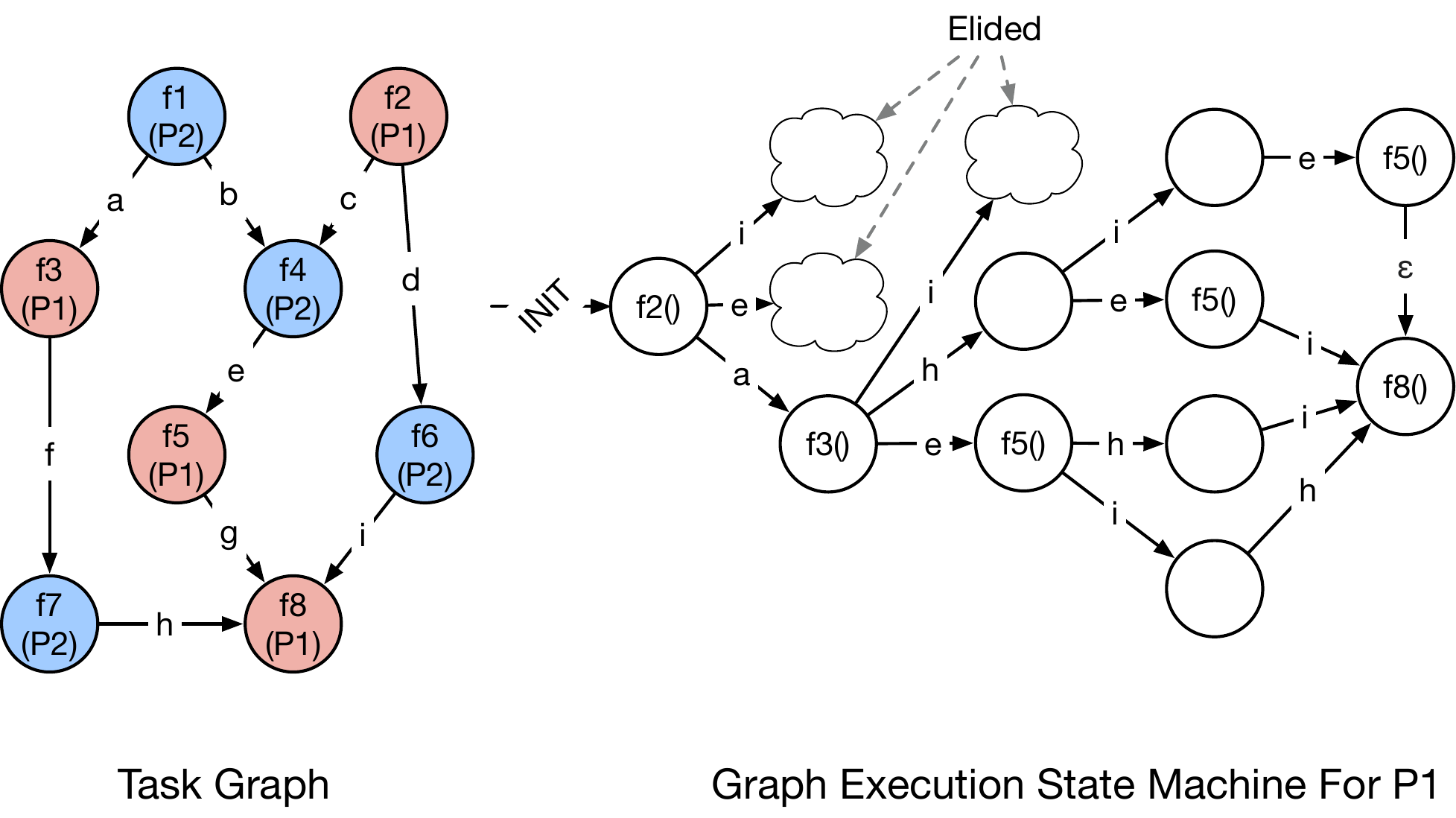}
\caption{Example task graph and a parallel message handling state machine.}
\label{fig:implicit-state-machine-viz}
\end{minipage}\hfill%
\begin{minipage}{0.33\textwidth}
\centering
\includegraphics[width=0.85\textwidth]{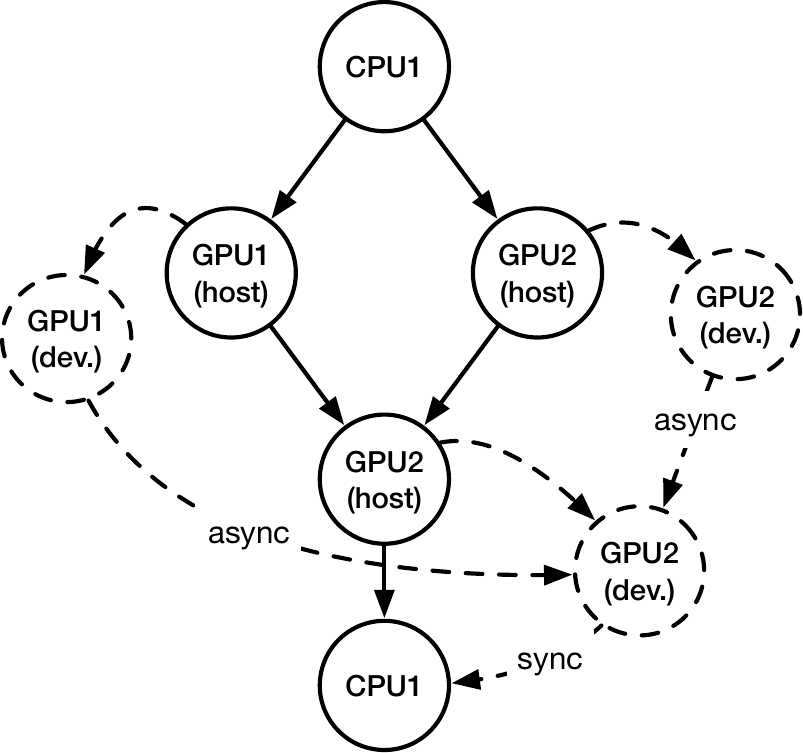}
\caption{Transformation for accelerators.
Dashed nodes and edges are added.}
\label{fig:gpu-graph-transform}
\end{minipage}
\end{figure*}

The goal of compilation is to specialize the runtime
system itself to the input task graph $G$, resulting in a set
of actors specialized to $G$~\cite{futamura}.
These specialized actors avoid expensive synchronization
structures (like events) and communicate with the minimum number
of cross-actor messages.
Constructing specialized actors involves building
a state machine for each actor that executes vertices in $G$
and transitions upon receiving messages from other actors.
To maximize performance, the state machine must exploit
all potential parallelism in $G$ by executing vertices
as soon as their dependencies are satisfied.
Topologically unordered vertices in $G$ may execute
in any order, and the completion messages for these vertices
may also arrive in any order.
The state machine handling all of these potential orderings
to execute vertices in $G$ as soon as possible would
contain a state for each prefix of every topological ordering
of $G$, as any one ordering may result in different
vertices being ready at any given point in time.
An example task graph and state machine 
handling all valid topological execution orderings 
is shown in \Cref{fig:implicit-state-machine-viz}.
Explicitly enumerating and generating code for such a state
machine in the syntactic style of \Cref{fig:simple-dag-actor-state-machine}
would be infeasible due to the factorially large number of
states\footnote{Managing this large state space is difficult for humans too,
who often tradeoff exploiting available parallelism to 
simplify the encoded state machine.}.
Instead, we separate the process into two phases, described
in \Cref{fig:compilation-algorithm}.
A compilation phase first pre-processes $G$ 
and constructs a set of actors.
Then, an interpretation phase executes the graph, where
each actor interprets a series of commands to execute operations. 
The factorially large state machine is encoded implicitly
through different dynamic configurations
of each actor's interpreter data structures.

\begin{algorithm}[t]
\caption{Task Graph Compilation Algorithm}
\label{fig:compilation-algorithm}
\footnotesize

\SetInd{0.5em}{0.75em}

\SetKwRepeat{StructActor}{class Worker$(r, (V, E))$}{}%
\SetKwProg{HandleMessage}{HandleMessage}{}{}
\SetKwProg{Worker}{class Worker}{}{}
\SetKwProg{Compile}{Compile}{}{}
\SetKwProg{Execute}{Execute}{}{}
\SetKwFunction{FSendMessage}{SendMessage}
\SetKwFunction{FExecute}{Execute}
\SetKwFunction{FRegisterActor}{RegisterActor}

\DontPrintSemicolon

\Compile{$(G)$}{
  $resources \gets$ all processors and memory channels used in $G$ \;
  \ForEach{$r \in resources$}{
    $V_w \gets$ nodes of $G$ running on $r$\;
    \tcc{An edge list pre-processed for O(1) lookup of edges. Maintains an atomic counter for each node in $V_w$'s incoming edges.}
    $E_w \gets$ in and out frontier of $V_w$ in $G$\;
    \FRegisterActor{$\text{Worker}(r, (V_w, E_w))$, $r$}\;
  }
}
\Worker{$(r, (V, E))$}{
\HandleMessage{$(id, data)$}{
  \Switch{$id$}{
    \Case{$INIT$}{
      $E \gets$ reset$(E)$\;
      \tcc{Start all ready to execute work.}
      \ForEach{$v \in V~|~\not\exists~(\_, v) \in E$}{
        \FSendMessage{$this$, $EXECUTE\_OP$, $v$}
      }
    }
    \Case{$COMPLETED\_EDGE$}{
      $(src, dst) \gets$ unpack$(data)$\;
      \tcc{Concretely computed by an atomic decrement to an indexing structure.}
      $E \gets E - (src, dst)$\;
      \If{$(\_, dst) \not\in E$}{
        \FSendMessage{this, $EXECUTE\_OP$, $dst$}
      }
    }
    \Case{$EXECUTE\_OP$}{
      $op \gets$ unpack$(data$)\;
      \tcc{Execute op on this actor's resources.}
      \FExecute{$op, r$}\;
      \ForEach{$(op, dst) \in E$}{
        \FSendMessage{owner$(dst)$, $COMPLETED\_EDGE$, $(op, dst)$}
      }
    }
  }}}
  
\Execute{$(G)$}{
  \ForEach{$w \in$ RegisteredWorkers$(G)$}{
    \FSendMessage{$w, INIT$}\;
  }
}
\end{algorithm}

The compilation step collects all processors and
data movement channels used by vertices of $G$.
For each resource $r$, we take the subgraph of $G$ that runs
on $r$ (referred to as $G_r$) 
and pre-process $G_r$ into an indexing structure
that enables constant-time lookup of edges.
Then, a set of counters is prepared for each vertex in $G_r$
that maintains the number of pending incoming dependencies for that vertex.
The combination of counter values for each vertex in $G_r$ corresponds
to a logical state in an actor state machine running on $r$.
Finally, the compilation step creates a worker actor for executing the corresponding subgraph associated with resource $r$.

Each worker actor is structured as an interpreter that processes
incoming messages and updates its state. 
Upon receiving a message to execute an operation, the worker executes
the corresponding operation
and then sends a message to the actors responsible
for running each dependent operation in $G$.
When receiving a message that an incoming edge has completed,
the worker decrements the corresponding counter and potentially
enqueues the target operation for execution.
Graph execution is initiated by sending each worker actor
an initialization message, which upon receiving, each worker 
enqueues all operations with no predecessors.

The worker actors themselves contain the minimal functionality for correctness to ensure low-overhead task graph execution.
Because $G$ is known and fixed, the
signaling of dependencies can be done without
any intermediate structures or extra messages, such as those
described in \Cref{sec:generic-task-actor-reduction}.
Furthermore, the actors eschew complex
concurrent data structures, and instead coordinate dependencies
within an actor using lock-free atomic decrements for each edge in $G$.
We show in \Cref{sec:evaluation} that this compilation strategy
improves overheads in the Realm runtime system by 1.7-5.3x.

\subsection{Optimizations for Accelerators}

%
To efficiently utilize asynchronous accelerators like GPUs, applications
must avoid blocking on the results of asynchronous operations
(i.e. \emph{run ahead})
and dispatching to hardware-supported synchronization whenever possible.
The dynamic nature of task-based models can make achieving these
two goals challenging.
Since dependencies between tasks are dynamic and unpredictable,
task-based systems often lift specialized accelerator dependence
structures into the dependence structure
modeled by the task-based system.
For example, the Realm~\cite{realm} runtime lifts CUDA
events into Realm events and triggers the lifted Realm 
events by polling the CUDA events in the background, missing 
opportunities for run ahead and performing more expensive synchronization.

We specialize the target task graph $G$ for
asynchronous accelerators by decoupling the
host-side of operations that launch asynchronous work 
from the device-side asynchronous work itself, as
visualized in \Cref{fig:gpu-graph-transform}.
%
For every node $n$ within $G$ that launches
accelerator work 
we add a new \emph{async} node $n_a$ representing the asynchronous work.
Then, for every outgoing edge $(n, d) \in G$, if $d$ has a corresponding
async node $d_a$, we add an async edge 
$(n_a, d_a)$.
If $d$ does not have a corresponding async node, then we
add a \emph{sync} edge $(n_a, d)$.
When executing $G$ using \Cref{fig:compilation-algorithm}, every asynchronous
operation records state representing its completion,
such as a CUDA event. 
Async nodes with async edges synchronize
against the corresponding state, such as predicating a CUDA 
kernel on a recorded CUDA event.
Sync edges are lowered by sending a message
to the destination actor when the source asynchronous operation completes.
This transformation enables significant run ahead 
and offloads synchronization to hardware-supported mechanisms whenever possible.
While presented in the context of accelerators, this procedure
could be used wherever the computation in a task is decoupled into separate
stages, such as when performing I/O.

\section{Lowering Implicitly Parallel Models}\label{sec:implicit-par}

We have demonstrated how task graphs in explicitly-parallel
task-based models can be compiled into actor programs that
execute with low overheads.
While explicitly-parallel models are useful for certain
applications, implicitly-parallel models provide
even larger productivity improvements by automatically
extracting parallelism from a sequentially expressed application.
We now discuss how to integrate our graph compilation infrastructure
into an implicitly-parallel task-based system, 
specifically within a \emph{tracing}~\cite{legion-tracing, auto-tracing} module.

Tracing memoizes the dependence analysis performed by an 
implicitly-parallel task-based runtime system.
Applications demarcate the start and end of a \emph{trace},
a repeatedly executed program fragment.
The runtime then memoizes the dependence analysis for the trace by
recording all computed dependencies between tasks in the trace.
On future invocations of the trace, the runtime executes tasks
by simply issuing them with the memoized dependencies (targeting an
interface like in \Cref{fig:task-based-runtime-api}).
To scale trace compilation and replay, Legion~\cite{legion} implements
tracing in a sharded (or control-replicated) 
manner~\cite{static-ctrl-repl, dyn-ctrl-repl}.
Each participating node $n$ memoizes
and optimizes only the subset of the global task graph that executes on $n$.
Then, on trace replay, nodes collaborate to enforce dependencies that
cross node boundaries.

\begin{figure}[t]
    \centering
    \includegraphics[width=.7\linewidth]{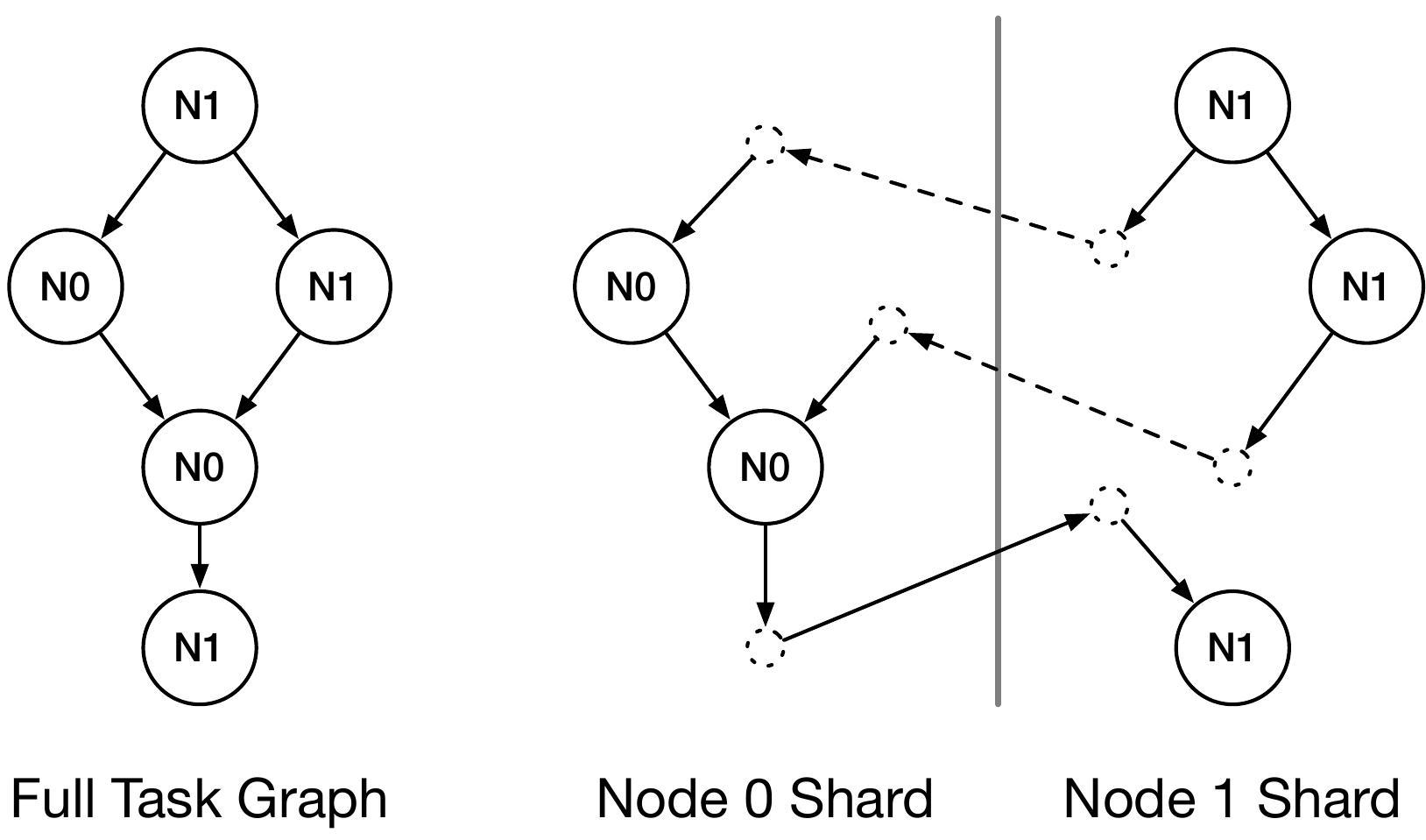}
    \caption{Example of task graph sharded onto two nodes.}
    \label{fig:sharded-graphs}
\end{figure}

We integrate our work into this sharded framework by representing
each node-local trace subset as a compiled graph.
We leverage the external pre-condition and post-condition vertices
of graph to coordinate the inter-node dependencies,
while intra-node dependencies are
expressed through direct edges, as visualized in \Cref{fig:sharded-graphs}.
Lowering traces onto compiled graphs further reduces 
the overhead imposed by the implicitly-parallel runtime system:
tracing itself removes dynamic analysis overheads and 
compiling the task graphs further reduces the overheads imposed
by the underlying explicitly-parallel runtime system.
We show in \Cref{sec:evaluation} that combining tracing with
task graph compilation yields significant improvements in 
strong-scaling performance.

This sharded approach is a tradeoff
between compilation time and execution performance.
As the size of the global task graph scales with the size of the
target machine, optimization and analysis of the entire graph scales similarly.
This cost can become prohibitive if non-linear optimizations are applied,
such as a transitive reduction to remove unnecessary edges.
Even representing the global task graph can require too much
memory and become infeasible at large scales.
On the other hand, sharding the global task graph into a compiled
graph per node relinquishes some performance, as dependencies visible
within a compiled graph can be optimized further than opaque graph
edges represented by external pre- and post-conditions.
We chose the sharded approach to maximize scalability
while also maximizing the potential for overhead reduction: intra-node
dependencies can often be satisfied with hardware-supported, light-weight communication
mechanisms, while inter-node dependencies are fundamentally limited by
the cost of a network message.

\section{Evaluation}\label{sec:evaluation}

\paragraph{Overview.}
We evaluate our work on micro-benchmarks and end-to-end applications
implemented within the Legion~\cite{legion} and Realm~\cite{realm} runtime systems.
Using the Task Bench~\cite{task-bench} framework, we first show that our work
significantly lowers the overheads imposed by both explicitly-parallel and implicitly-parallel
task-based models, and enables explicitly-parallel models to
support fine-grained computations with competitive performance to
low-level actor frameworks like MPI~\cite{mpi} and Charm++~\cite{charm++}.
We then show that our work improves strong-scaling of end-to-end
applications developed in implicitly-parallel task-based systems.

\paragraph{Experimental Setup.}
We ran all experiments on an NVIDIA DGX H100 supercomputer, where each
node contains 8 H100 80GB GPUs and a 112 core Intel Xeon Platinum.
%
Nodes are connected with Infiniband.
We configure Legion and Realm with the GASNet-EX~\cite{gasnet-lcpc18} networking module.
We compare against Open MPI 4.1.7, Charm++ 6.9.0, StarPU 1.4.7,
Ray 2.47.1, and run all applications with CUDA 12.4.1.

\subsection{Measuring Overheads With Task Bench}\label{sec:eval-task-bench}

\begin{figure}
\centering
\includegraphics[width=0.7\linewidth]{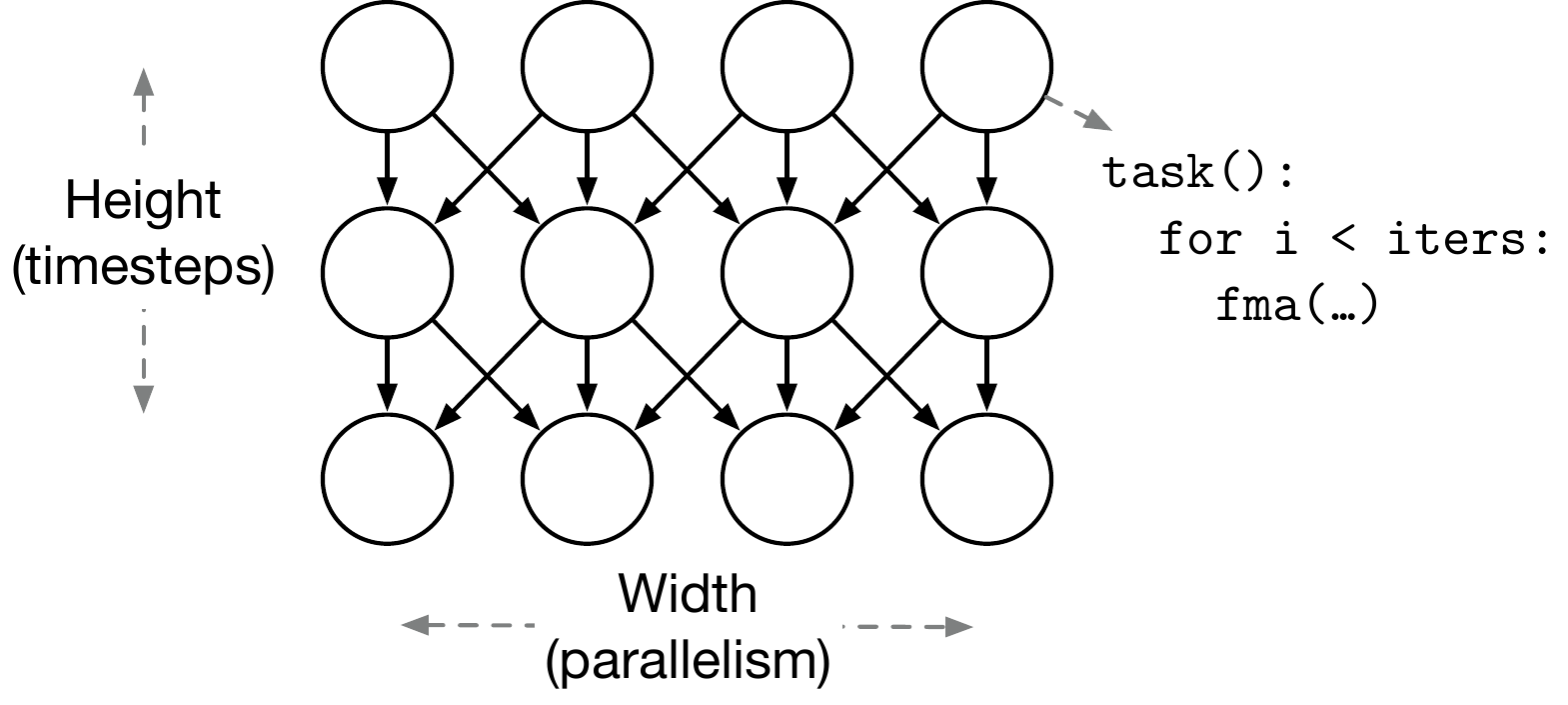}
\caption{Task Bench benchmark structure.}
\label{fig:task-bench-graph-example}
\end{figure}

\begin{figure}
\centering
\includegraphics[width=0.9\linewidth]{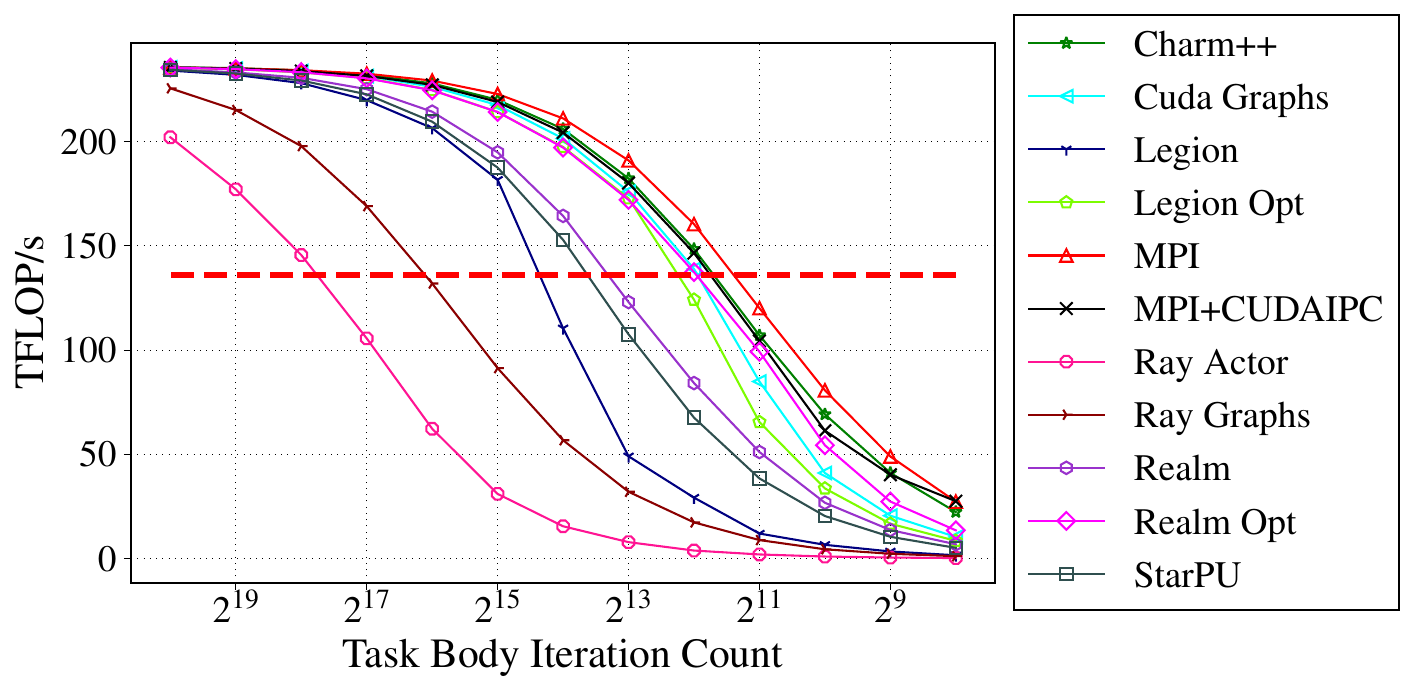}
\caption{FLOPs achieved on 1 node by each system on a stencil Task Bench graph with width 8.}
\label{fig:stencil-1-node-flops}
\end{figure}

\begin{figure*}
\begin{subfigure}[b]{0.49\textwidth}
\centering
\includegraphics[width=0.9\textwidth]{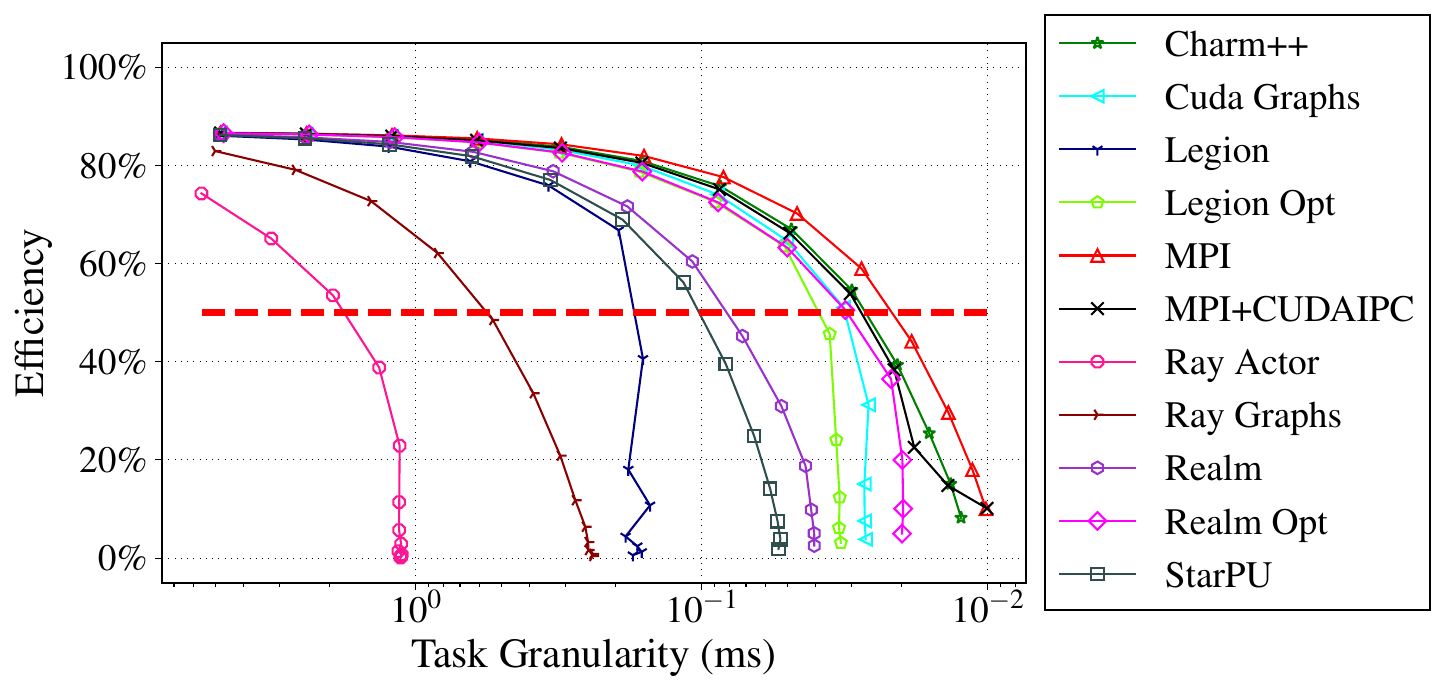}
\caption{1 node (8 GPUs), width 8}
\label{fig:tb-1-node}
\end{subfigure}\hfill
\begin{subfigure}[b]{0.49\textwidth}
\centering
\includegraphics[width=0.9\textwidth]{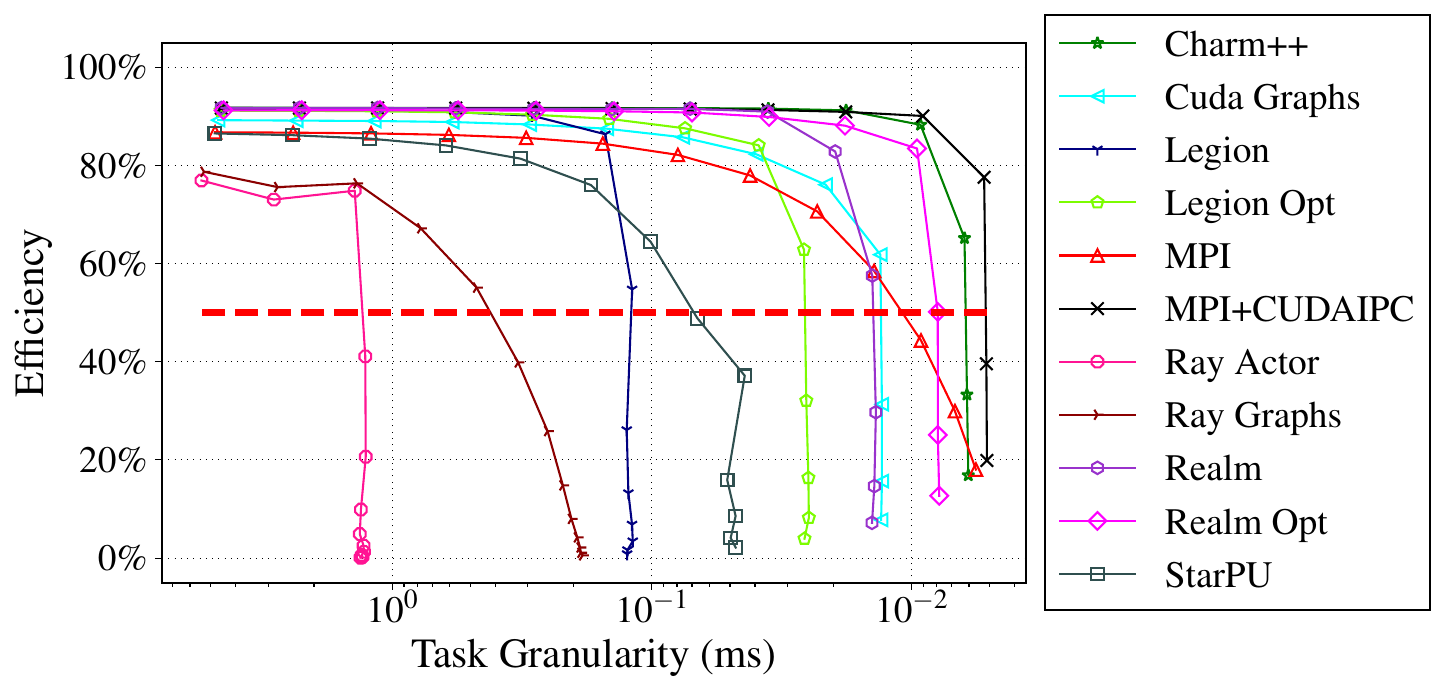}
\caption{1 node (8 GPUs), width 32}
\label{fig:tb-4g-1-node}
\end{subfigure}

\begin{subfigure}[b]{0.49\textwidth}
\centering
\includegraphics[width=0.9\textwidth]{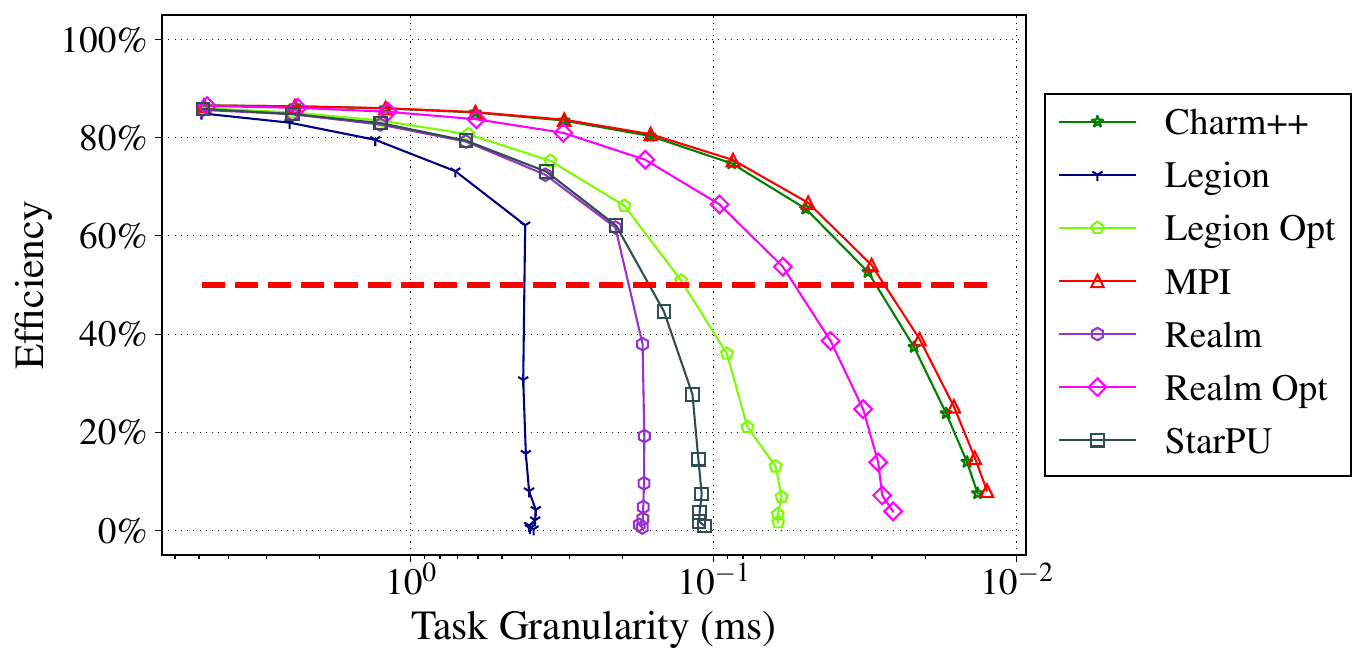}
\caption{4 nodes (32 GPUs), width 32}
\label{fig:tb-4-node}
\end{subfigure}\hfill
\begin{subfigure}[b]{0.49\textwidth}
\centering
\includegraphics[width=0.9\textwidth]{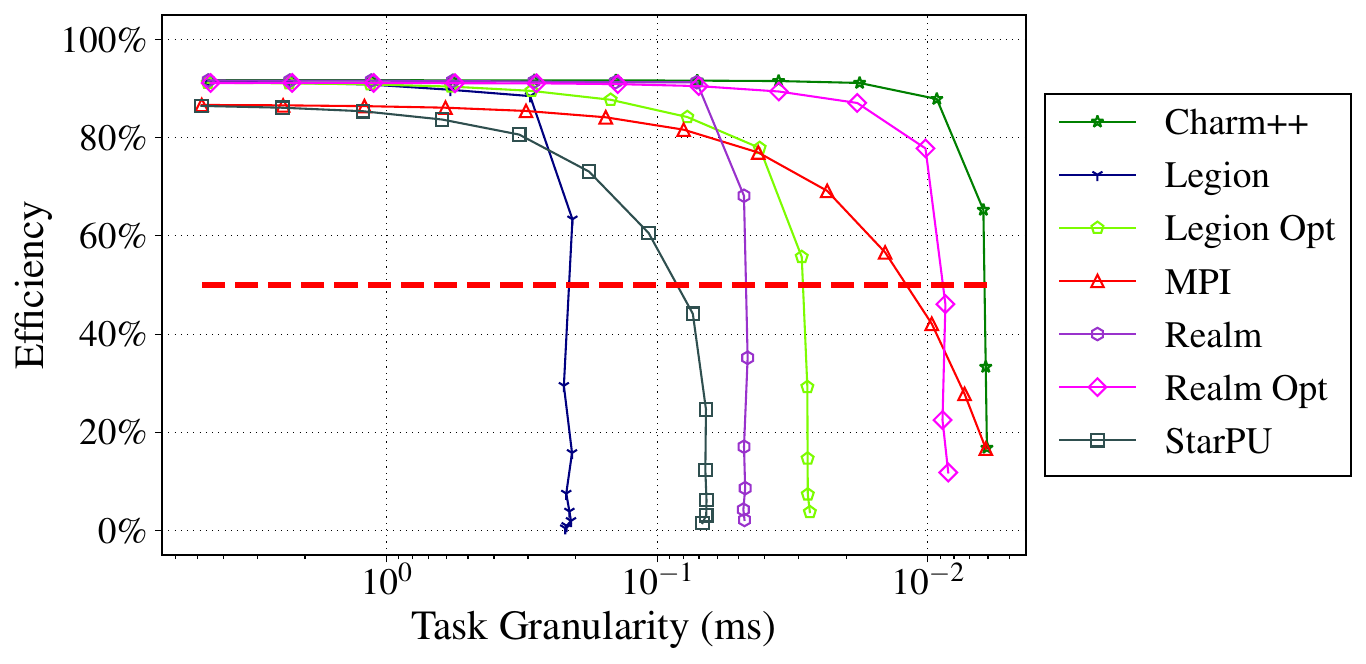}
\caption{4 nodes (32 GPUs), width 128}
\label{fig:tb-4g-4-node}
\end{subfigure}

\caption{Task Bench minimum effective task granularity (METG) curves of different systems on stencil task graphs.}
\label{fig:task-bench-metg}
\end{figure*}

Task Bench~\cite{task-bench} is a framework for comparing
runtime system overheads.
%
Task Bench defines a graph of tasks (atomic, coarse grained work items)
in a generic interface to be implemented by each benchmark system.
The task graph is a two-dimensional grid with width corresponding to
available parallelism, height corresponding to number of timesteps to execute,
and dependencies that relate tasks from timestep $t - 1$ to timestep $t$,
as shown in \Cref{fig:task-bench-graph-example}.
A Task Bench experiment fixes a graph structure, and 
varies only the amount of computation performed at each vertex.
\Cref{fig:stencil-1-node-flops} shows the FLOPs achieved by various systems
on a single node (8 GPUs) running on a Task Bench graph with a stencil dependence
structure (shown in \Cref{fig:task-bench-graph-example}) and width of 8.
Each Task Bench task launches a CUDA kernel that performs floating point operations
in a loop with the number of iterations controlled by the x-axis of the graph.
At high iteration counts (long task run time), most systems achieve close to
peak (non-tensor-core) FLOPS available on a DGX H100 (272 TFLOPS), but as the 
number of iterations decreases,  various overheads decrease the
total FLOPS achieved by each system.

To evaluate overhead, Task Bench uses these performance curves
to define a metric called \emph{minimum effective task granularity} (METG).
The METG(50) is the smallest task granularity where a system achieves at least
50\% of the peak FLOPS, quantifying the task granularity
at which overheads dominate the execution time.
METG is superior to several common alternative metrics for measuring runtime system efficiency: weak-scaling hides arbitrary
overhead if the problem sizes are too large, strong-scaling does not separate
changing application costs (e.g. increased communication) from runtime costs, 
and tasks-per-second fails to consider the amount of useful application work performed.
See \cite{task-bench} for a complete discussion of METG.
%
%
For non-task-based systems, METG encapsulates the amount of application work
required to offset operations like messaging and synchronization.

We compare versions of Legion and Realm optimized with the techniques in this
paper (called Legion Opt and Realm Opt) against the well-known HPC actor systems
Charm++~\cite{charm++} and MPI~\cite{mpi},
the implicitly-parallel tasking system StarPU~\cite{starpu}, and
the popular task and actor system from the ML community, Ray~\cite{ray}.
On a single node, we also compare against CUDA Graphs
and an MPI implementation that uses CUDA IPC Events to lower GPU synchronization
overheads within a single node.
When possible, implementations for each system were taken as-is
 from the previously published Task Bench implementations,
or adapted to utilize GPUs.

\Cref{fig:tb-1-node} presents curves that plot task granularity against the
efficiency achieved by various systems on a single node with a stencil
task graph of width 8.
These curves are derived from \Cref{fig:stencil-1-node-flops},
where the x-axis now corresponds to the runtime in milliseconds of each
Task Bench task, instead of the number of kernel iterations.
We see three distinct groups of systems, in order
of increasing METG: actor and compiled task-based systems,
standard task-based systems, and Ray.
This configuration's METG is mostly determined by
how fast each system can issue CUDA kernels.
The actor and compiled task-based systems achieve METG(50)s between
22us-39us (MPI and Legion Opt, respectively).
%
These systems are also competitive with CUDA Graphs; since we expect
CUDA graphs to be an efficient way of launching kernels, this demonstrates
we are achieving high absolute efficiency.
We are separately interested in why these systems can outperform CUDA
Graphs, but that is beyond the scope of this paper.
%
The standard task-based systems (Legion, Realm and StarPU) accumulate overheads
from numerous sources, achieving METG(50)s between 83us-173us.
The final system is Ray, where we report results using Ray's
actors as well as Ray's graph compiler~\cite{ray-dag}.
We initially developed an implementation purely using Ray's tasking interface,
but found the performance to be too poor to easily visualize with the other systems; each Ray task created a
new process, causing re-initialization of structures like the CUDA runtime,
requiring task granularity of at least hundreds of milliseconds.
Ray Actors achieve a METG(50) of 1.8ms, which is improved by compilation
to 570us, both larger than HPC tasking systems.
These overheads are likely due to Ray supporting features 
such as resilience and elasticity.

\begin{figure*}
\begin{subfigure}[b]{0.24\textwidth}
\centering
\includegraphics[width=\textwidth]{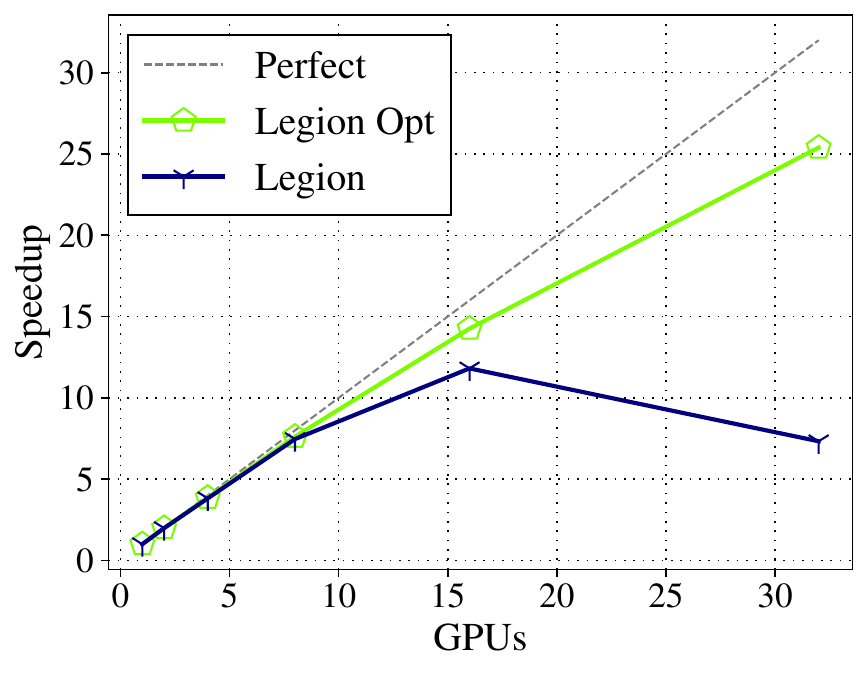}
\caption{2D Stencil}
\label{fig:stencil-strong-scaling}
\end{subfigure}\hfill
\begin{subfigure}[b]{0.24\textwidth}
\centering
\includegraphics[width=\textwidth]{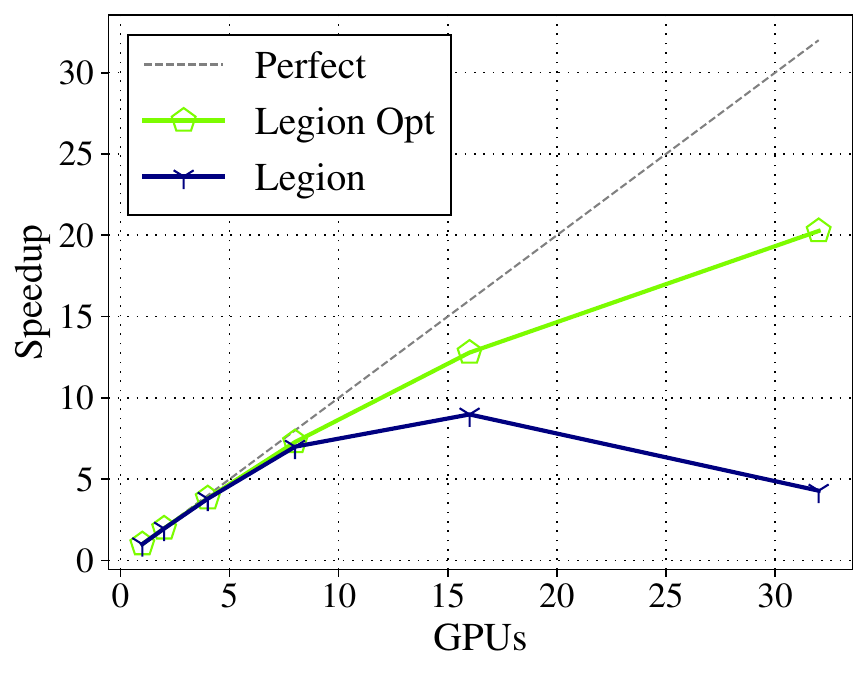}
\caption{MiniAero}
\label{fig:miniaero-strong-scaling}
\end{subfigure}\hfill
\begin{subfigure}[b]{0.24\textwidth}
\centering
\includegraphics[width=\textwidth]{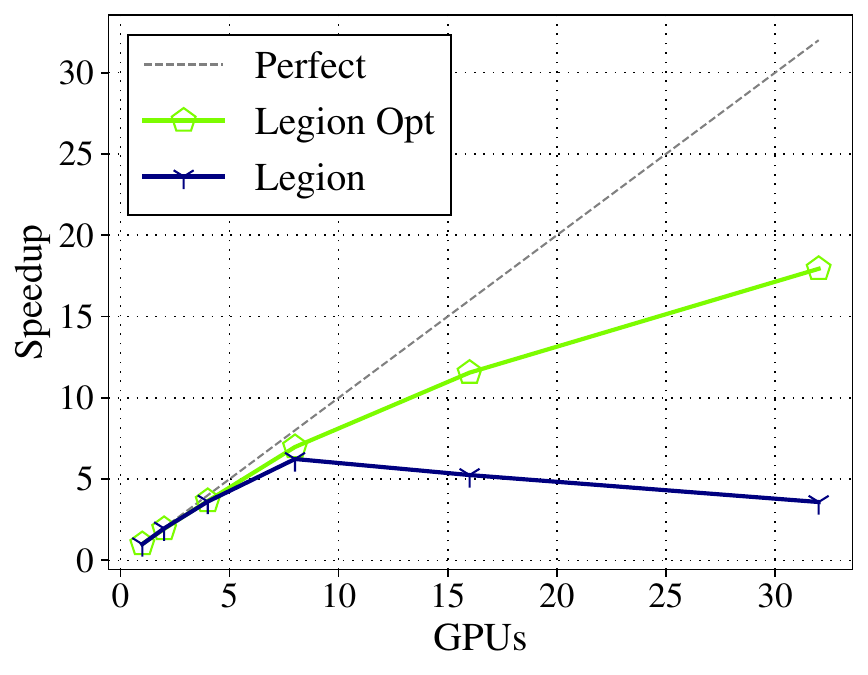}
\caption{PENNANT}
\label{fig:pennant-strong-scaling}
\end{subfigure}\hfill
\begin{subfigure}[b]{0.24\textwidth}
\centering
\includegraphics[width=\textwidth]{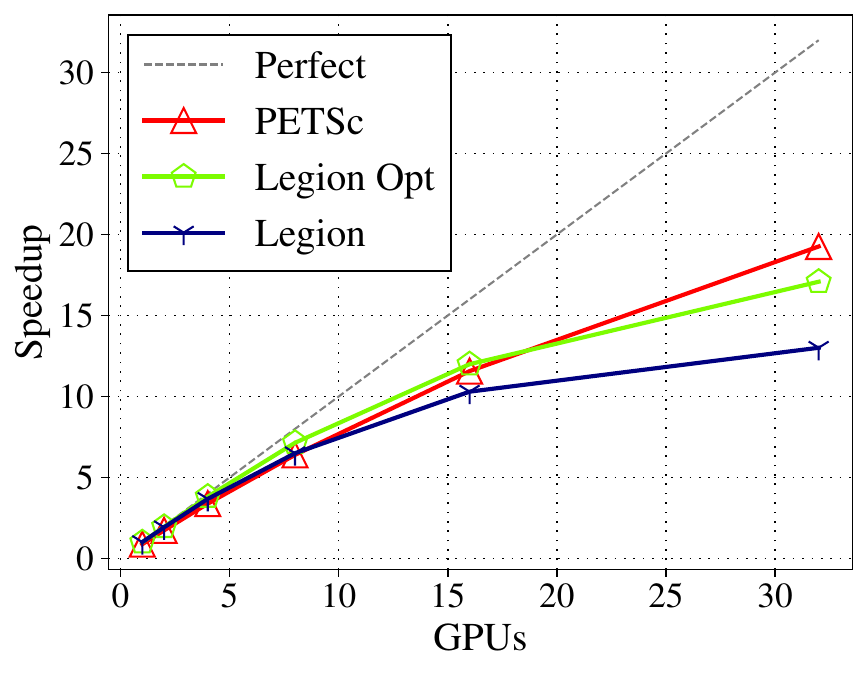}
\caption{Conjugate Gradient}
\label{fig:cg-strong-scaling}
\end{subfigure}

\caption{Strong scaling performance of end-to-end Legion applications (higher is better).}
\label{fig:strong-scaling}
\end{figure*}




\Cref{fig:tb-4g-1-node} contains single-node results
with a graph of width 32, exposing 4-way task-parallelism
on each GPU and offering an opportunity to hide latency.
With task parallelism, the METG(50) of HPC systems
falls into the single microseconds, where MPI+CUDAIPC, Charm++ and
Realm Opt achieve METG(50)s of 5.1us, 6.1us and 7.9us respectively.
Other systems fall off earlier for different reasons.
The vanilla MPI implementation is written in a bulk-synchronous style,
unable to perform fine-grained interleaving of CUDA kernels that
the MPI+CUDAIPC implementation can.
%
%
Legion Opt achieves a METG(50) of 25us (a 4.6x improvement over Legion)
despite Realm opt achieving a METG(50) of 7.9us.
This overhead originates from within task execution itself:
for correctness reasons, Legion tasks always query the runtime system for pointers
to the task's data in case different mapping decisions for a task's data have been made from one iteration to the next.
These queries took roughly 20us in aggregate to complete, placing a floor
on the smallest tasks that Legion can execute.
In contrast, the Task Bench implementations in lower level systems like 
Realm (or Charm++ and MPI) preallocate all necessary data,
and pass direct pointers into task invocations to avoid overheads.

We now move to multiple nodes, where \Cref{fig:tb-4-node} contains results 
on 4 nodes using a 32-wide graph (1 task per GPU).
Like \Cref{fig:tb-1-node}, this configuration is latency constrained,
without parallelism to hide communication costs.
Charm++ and MPI perform the best, achieving METG(50)s of 
of 29us and 27us respectively.
Realm Opt achieves a METG(50) of 54us, 3.5x better than standard Realm.
The performance difference between Realm Opt and Charm++/MPI arises from the
tradeoff discussed in \Cref{sec:implicit-par}:
the Realm Task Bench implementation is developed in a sharded style
for scalability, where each node constructs a local graph and connects
to other nodes with Realm's standard dependence infrastructure.
While still making a single network round-trip, the dependence infrastructure 
has higher overhead than a Charm++/MPI message handler.
Legion Opt achieves an METG(50) of 125us, 2.5x better than standard Legion.
While Ray is a distributed runtime system, we did not evaluate it due to
the lack of competitiveness on a single node.

The final Task Bench experiment is in \Cref{fig:tb-4g-4-node}, a 4-node
configuration with 4-way task parallelism on each GPU.
The results are similar to \Cref{fig:tb-4g-1-node}, where Charm++ achieves a METG(50) of
6.1us and Realm Opt achieves a METG(50) of 8.8us (a 5.2x improvement over standard Realm);
the bulk-synchronous MPI implementation degrades in performance earlier.
With some task-parallelism to exploit, Realm Opt is able to close the performance difference
experienced in \Cref{fig:tb-4-node}.
Legion Opt supports a similar METG(50) as on a single node (28us), running
into the same intra-task overheads, but achieving a 7.1x improvement over standard Legion.

Our experiments show that our work dramatically lowers the
overheads imposed by both explicitly-parallel and implicitly-parallel task-based systems,
improving the METG(50) of both Legion and Realm by between 3.3x-7.1x and 1.77-5.3x respectively.
By transforming critical subsets of task-based programs into actors, we recover a significant amount of the performance
difference with native actor programs, and deliver competitive runtime system overheads not previously achieved by existing
task-based systems.

\subsection{Strong Scaling Implicit Parallelism}

%
We now now demonstrate these reduced overheads yield end-to-end strong-scaling
improvements for applications developed in 
implicitly-parallel task-based systems.
In each application, Legion is already performing tracing to eliminate dynamic
dependence analysis overheads --- the remaining difference is the efficiency
of the traced task graph's execution.
Each benchmark application has been heavily optimized separately
from this work, and many have appeared in existing 
publications~\cite{static-ctrl-repl, dyn-ctrl-repl, legion-tracing}.
We also note that while strong-scaling studies tend to present
very large problem sizes (often not even fitting into a single GPU memory), our experiments
instead start scaling at a modest problem size, filling roughly half the memory of
just one H100 GPU.
The time spent in graph compilation was negligible, less than 50ms
on every node.

\paragraph{Stencil}

The smallest application is a stencil benchmark from the
Parallel Research Kernels~\cite{prk}, with results in \Cref{fig:stencil-strong-scaling}.
The benchmark performs a radius-2 star-shaped stencil over a two-dimensional
grid.
The benchmark has no task parallelism, and is thus sensitive
to any latencies when launching kernels or performing the halo exchange
at grid boundaries.
As a result, Legion Opt is able to continue improving performance after 16 GPUs while
standard Legion falls over, resulting in an improvement of 3.4x at 32 GPUs.

\paragraph{MiniAero}

MiniAero is a 3D unstructured mesh proxy application from the Mantevo suite~\cite{mantevo}
that implements an explicit solver for the compressible Navier-Stokes equations,
with results in \Cref{fig:miniaero-strong-scaling}.
Unlike Stencil, MiniAero has many opportunities to exploit task parallelism, benefiting from lower
overheads to enable more utilization of the GPUs.
Similarly to the performance of Stencil, Legion Opt continues scaling
after 16 GPUs, while standard Legion falls off after 16, achieving a
4.7x improvement at 32 GPUs.
We attempted to compare against the reference implementation, 
but found that it depended on unsupported
versions of Kokkos, and did not run when ported to a newer version.

\paragraph{PENNANT}

PENNANT is a 2D unstructured mesh proxy application simulating
Lagrangian hydrodynamics~\cite{pennant}, with results in
\Cref{fig:pennant-strong-scaling}.
%
%
The PENNANT main loop additionally contains some all-reduce operations
that cannot be hidden by parallel work, and thus are exposed and affect
the total speedup achievable.
As part of graph compilation, we pre-plan and optimize copies
present in the task graph, lowering the latency of small copy operations, which
improves the all-reduce performance.
The standard Legion implementation falls over at 8 GPUs, while Legion Opt continually
delivers speedup up to 32 GPUs, achieving a 5.0x improvement.
The reference PENNANT GPU implementation only runs on a single GPU, and employs
optimizations that make it not directly comparable to a multi-GPU implementation.

\paragraph{Conjugate Gradient}

Our final benchmark is a conjugate-gradient (CG) solver for a 1-D Poisson problem,
with scaling shown in \Cref{fig:cg-strong-scaling}.
We additionally compare against a baseline implementation developed
using PETSc~\cite{petsc}, an industry-standard sparse linear algebra library developed with MPI.
Legion Opt, standard Legion and PETSc all strong scale well, with
PETSc doing the best, and Legion Opt improving upon the scalability of Legion.
Legion Opt achieves 88\% of PETSc's performance at 32 GPUs, while standard Legion
only achieves 67\%.
The conjugate gradient application has very little task-parallelism, similar to \Cref{fig:tb-4-node}.
In the 4-node Task Bench configuration without task-parallelism, Legion Opt's METG(50) was
125us, which the average task duration of CG at 32 GPUs approaches, corresponding to the
beginnings of the performance difference against PETSc.

\section{Related Work}

\paragraph{Actors}

Actor-based models are used extensively and underlie significant
amounts of distributed software.
Many runtime systems are implemented as actors that
communicate through low-level networking layers~\cite{gasnet-lcpc18, ucx}.
Numerous frameworks have been developed~\cite{charm, charm++, mpi, ray, cpp-native-actors, orleans, erlang, houck-actors}
that embed actor functionality within a host language and provide convenient language
features for communication between actors.
Within these frameworks, languages have been proposed to simplify the state management
involved in developing large scale actor programs~\cite{structured-dagger}.
Finally, actor-based models have received significant
formal study~\cite{agha-actors, actors-grief}, and built on work
developing process calculi~\cite{milner-ccs}.

\paragraph{Tasks}

Many task-based systems have been developed by the
HPC~\cite{legion, regent, parsec, starpu, cudastf, realm},
cloud computing~\cite{spark}, and ML~\cite{ray, tensorflow, pathways} communities.
Each system was designed with different priorities and driving applications
but at their core allow for the programmer to execute a parallel computation graph.
The methods to describe and construct the underlying graph differ across systems, 
such as through automatic extraction, algebraic description or explicit user construction.
The smallest task sizes efficiently supported by existing systems
can differ by multiple orders of magnitude~\cite{task-bench}.
%

Memoization and compilation are standard techniques used in task-based
models to lower overheads in repeatedly executed computations.
The Legion runtime system uses tracing~\cite{legion-tracing} to memoize
the costs of dependence analysis.
%
Recent work by Yadav et al.~\cite{auto-tracing} shows how implicitly-parallel
runtime systems can automatically find traces to memoize, accelerating
task-based programs without user intervention.
Compiling traced task graphs into low-level actors is the final step of
this memoization pipeline, removing the underlying explicitly-parallel 
runtime system's overhead from iterative execution.
On a single node, CUDA Graphs~\cite{cuda-graphs} are used extensively
to lower the overheads of executing repeated graphs of CUDA kernels,
and are used as graph execution targets by single-node tasking runtimes
like CUDASTF~\cite{cudastf}.

\paragraph{Unifying Actors and Tasks}

Given the dichotomy between actor-based and task-based
programming models, researchers have introduced programming
models that attempt to provide support for both actors and tasks~\cite{ray, sarkar-actor-task-unify}.
%
%
This tension has been evinced within the Ray programming system~\cite{ray}, which was
initially conceived as a task-based programming model.
The overheads from the implementation of tasking were large enough that
actors were introduced into the language to lower overheads of certain computational patterns.
The actor model allowed Ray to lower runtime overheads, but exposes end-users to the difficulties
of state management discussed in \Cref{sec:specialized-actors}.
Our work shows how these overheads may be avoided while retaining the productivity benefits
of task-based programming models.

\IGNORE{
\TODO{intro into the ray stuff, cite some more projects that offer both models.}
Perhaps nowhere is the tension between tasks and actors better evinced than in the Ray programming system~\cite{ray}. 
The Ray runtime was initially conceived with a task-based programming model.
However, the overheads associated with their implementation were sufficiently large that the developers chose to rebrand Ray and implement an entirely new programming model based on actors.
The actor model allowed Ray to minimize runtime overheads, but punted the problem of productivity back up to their users.
Our results demonstrate that this was a wholly unnecessary course of action and that the same performance could have been achieved with their original task-based programming model without the extra development cost and incurred productivity burden to their users.
}

\paragraph{Message and Procedure Duality}

Our work was inspired by the duality between message-based and
procedure-based operating systems presented by Lauer et al.~\cite{os-duality}.
Lauer's work spurred further debate about whether
threading-based or event-driven architectures were the right choice
for productivity and performance~\cite{ousterhout-threads, brewer-events, haller-events}.
We take the position that in distributed programming, actor-based
and task-based models occupy different ends of a productivity and
performance tradeoff space, and demonstrate a technique to recover
actor performance in task-based programs without sacrificing productivity.
Follow on work explored similar dualities in other domains,
such as in fault-tolerance~\cite{ft-duality}.

\section{Conclusion}

In this work, we described a duality between task-based and actor-based
programming models, and explored a trade-off space between performance
and productivity defined by the two models.
Through this duality, we develop a compilation and execution strategy
for task-based programming models that lower task graphs onto a set
of specialized actors, greatly reducing the overheads that task-based
programming models impose.
We show through implementations within the Realm and Legion runtime
systems that our approach can make task-based programming systems
offer overheads competitive with actor-based systems, and can
significantly improve the strong-scaling of implicitly-parallel
programming systems.
Our work defines a new point on the Pareto frontier between 
performance and programmability in distributed programming models,
by retaining the programmability of task-based models while 
greatly increasing performance.

\section{Acknowledgments}

We thank Elliot Slaughter for help with Regent and Task Bench.
We thank Laxmikant Kale for help with Charm++, Samuel Thibault for
help with StarPU and Melih Elibol for help with Ray.
We thank Artem Priakhin, Cory Perry and Wei Wu for support with Realm.
We thank Benjamin Driscoll, Matthew Sotoudeh  and David Zhang for help
with the theoretical formulations of the reductions.
We thank (in no particular order) Melih Elibol, Scott Kovach, Shiv Sundram,
Manya Bansal, Chris Gyurgyik, AJ Root and Bobby Yan for feedback on this draft.
We thank the developers of the MagicTrace profiling tool, which helped
start and guide the investigation into this work.
A portion of this work was done while Rohan Yadav was at NVIDIA Research.

\bibliographystyle{ACM-Reference-Format}
\bibliography{main.bib}

\end{document}